\documentclass[showpacs,prl,onecolumn,aps,superscriptaddress,preprintnumbers,nofootinbib]{revtex4}
\usepackage[T1]{fontenc}
\usepackage[latin9]{inputenc}
\usepackage{amsmath,amssymb}
\usepackage{epsfig}
\usepackage{graphicx}
\usepackage{amsmath}
\usepackage{amsfonts}
\def\slashchar#1{\setbox0=\hbox{$#1$}     		
   \dimen0=\wd0                                 	
   \setbox1=\hbox{/} \dimen1=\wd1               	
   \ifdim\dimen0>\dimen1                        	
      \rlap{\hbox to \dimen0{\hfil/\hfil}}      	
      #1                                        	
   \else                                        	
      \rlap{\hbox to \dimen1{\hfil$#1$\hfil}}   	
      /                                         	
   \fi}

\renewcommand{\vec}{\boldsymbol}
\newcommand{\beq}{\begin{equation}}
\newcommand{\eeq}{\end{equation}}
\newcommand{\bea}{\begin{eqnarray}}
\newcommand{\eea}{\end{eqnarray}}
\newcommand{\ba}{\begin{array}}
\newcommand{\ea}{\end{array}}

\def\eq#1{{Eq.~(\ref{#1})}}
\def\fig#1{{Fig.~\ref{#1}}}

\newcommand{\as}{\alpha_S}

\newcommand{\Lb}{\left(}
\newcommand{\Rb}{\right)}
\def\pom{{I\!\!P}}
\begin{document}
\title{CGC/saturation approach: soft interaction at the LHC energies }
\author{E. ~Gotsman}
\email{gotsman@post.tau.ac.il}
\affiliation{Department of Particle Physics, School of Physics and Astronomy,
Raymond and Beverly Sackler
 Faculty of Exact Science, Tel Aviv University, Tel Aviv, 69978, Israel}
 \author{E.~ Levin}
 \email{leving@post.tau.ac.il, eugeny.levin@usm.cl}
 \affiliation{Department of Particle Physics, School of Physics and Astronomy,
Raymond and Beverly Sackler
 Faculty of Exact Science, Tel Aviv University, Tel Aviv, 69978, Israel}
 \affiliation{Departemento de F\'isica, Universidad T\'ecnica Federico
 Santa Mar\'ia, and Centro Cient\'ifico-\\
Tecnol\'ogico de Valpara\'iso, Avda. Espana 1680, Casilla 110-V,
 Valpara\'iso, Chile}  
 \author{I. Potashnikova}
 \email{irina.potashnikova@usm.cl}
  \affiliation{Departemento de F\'isica, Universidad T\'ecnica Federico
 Santa Mar\'ia, and Centro Cient\'ifico-\\
Tecnol\'ogico de Valpara\'iso, Avda. Espana 1680, Casilla 110-V,
 Valpara\'iso, Chile}  
\date{\today}

\keywords{BFKL Pomeron, soft interaction, CGC/saturation approach, 
Pomeron structure}
\pacs{ 12.38.-t,24.85.+p,25.75.-q}



\begin{abstract} In this paper we demonstrate that our model which is 
based
 on the CGC/saturation approach, is able to describe the soft interaction 
collisions including   the new TOTEM prelimenary data at 13 TeV. We 
believe that this   strengthens
 the  argument that the CGC/saturation approach is  the  only viable 
candidate for an  effective theory for high energy QCD.

\end{abstract}
 \preprint{TAUP-3029/17}

\maketitle


{\it Introduction:}\\
In our recent paper\cite{GLP} we have constructed a model, which allows us
 to discuss
  soft and  hard processes on the same footing, it is based on
 the CGC/saturation approach(see Ref.\cite{KOLEB} for a review) and on
 our previous attempts to build such a model
\cite{GLMNI,GLM2CH,GLMINCL,GLMCOR,GLMSP,GLMACOR,GOLE,GOLELAST}.
  
  In the model which we  proposed in Ref.\cite{GLP}, we  successfully
describe the 
DIS data from HERA, the total, inelastic, elastic and
 diffractive cross sections, the $t$-dependence of
 these cross sections, as well as the inclusive production and rapidity 
and angular
 correlations in a wide range of energies, including that of the LHC.
 
   Since the main feature, that we discuss in Ref.\cite{GLP}, are
 the angular correlations, our progress in describing the soft interaction
 data went unnoticed.  The representatives of the TOTEM collaboration,
reporting on their new
prelimenary results for energies W = 2.76 TeV \cite{Csorgo} and W = 13 TeV 
\cite{TOT},  neglected to mention
 our model as  one that provides a good description of their data 
(see Ref.\cite{TOT}).

  The goal of this letter is to draw the  attention of the high energy 
community
 that our approach  is successful in  describing the entire set of 
data for high
 energy soft scattering, including the  new experimental data from 
the LHC.\\
 
   
   {\it The model: theoretical input from  the CGC/saturation approach.}\\
Our model  incorporates two
 ingredients: the achievements of the CGC/saturation approach and the pure  
phenomenological
 treatment of the long distance non-perturbative  physics, necessary, due 
to the lack of
 the theoretical understanding of confinement of quark and gluons.
 
  For completeness of presentation we initially include a 
review of our approach, more details of which can be found in
\cite{GLP}.
 
   For  the effective theory for QCD at high energies we 
 have  two different formulations:  the CGC/saturation approach
 \cite{MV,MUCD,BK,JIMWLK}, and the BFKL Pomeron calculus 
\cite{BFKL,LI,GLR,GLR1,MUQI,MUPA,BART,BRN,KOLE,LELU,LMP,AKLL,AKLL1,LEPP}. 
  In Ref.\cite{AKLL1} it was shown, that 
these
 two approaches are equivalent for
\beq \label{MOD1}
Y \,\leq\,\frac{2}{\Delta_{\mbox{\tiny BFKL}}}\,\ln\Lb
 \frac{1}{\Delta^2_{\mbox
{\tiny BFKL}}}\Rb
\eeq
where $\Delta_{\mbox{\tiny BFKL}}$ denotes the intercept of the BFKL 
 Pomeron. In our model $ \Delta_{\mbox{\tiny BFKL}}\,
\approx\,0.2 - 0.25$    leading to $Y_{max} = 20 - 30$, which covers
 all collider energies.
 
  Bearing this equivalence in mind,    in constructing our model we rely
 on the BFKL Pomeron calculus, as
 the relation to diffractive physics  and soft processes,
 is more transparent  in this approach.

\begin{figure}[ht]
    \centering
  \leavevmode
      \includegraphics[width=14cm]{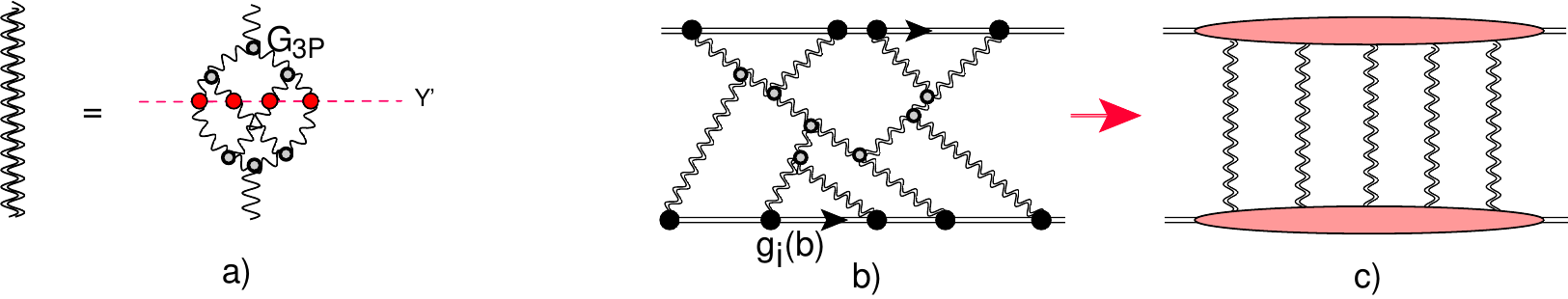}

      \caption{\protect\fig{amp}-a shows the set of  diagrams in the
 BFKL Pomeron calculus that produce the resulting (dressed) Green
 function of the Pomeron in the framework of high energy QCD. The red blobs
 denote the amplitude for the dipole-dipole interaction at low energy
 In \protect \fig{amp}-b the net diagrams,    which   include
 the interaction of the BFKL Pomerons with colliding hadrons, are shown.
 The sum of the diagrams after integration
 over positions of $G_{3 \pom}$ in rapidity, reduces to   \protect\fig{amp}-c.}
\label{amp}
   \end{figure}

  The main ingredient, that we need to find, is the resulting (dressed)
 BFKL Pomeron Green function, which 
 can be calculated using $t$-channel unitarity constraints:
\beq \label{GFMPSI}
 G^{\mbox{\tiny dressed}}_\pom\Lb Y, r, R; b \Rb  =\int \prod_{i =1} d^2 r_i\, d^2 b_i\, d^2 r'_i\, d^2 b'_i\,
 N\Lb Y- Y', r, \{\vec{r}_i,\vec{b} - \vec{b}_i\}\Rb
\fbox{$A^{\rm BA}_{\mbox{\tiny
 dipole-dipole}}
\Lb r_i, r'_i, \vec{b}_i - \vec{b'}_i\Rb$}\,
N\Lb  Y', R, \{\vec{ r}'_i,\vec{b}'_i\}\Rb
\eeq
where  $ N\Lb Y- Y', r, \{ \vec{r}_i,\vec{b} - \vec{b}_i\}\Rb$ denotes 
the amplitude
  for the  production in the  $t$-channel  of  the set of dipoles 
with $Y=Y'$
 and with the size $r_i$,
at the impact parameters $b_i$.
$A^{BA}_{\mbox{\tiny dipole-dipole}}$ denotes the dipole-dipole
 scattering amplitude in the Born approximation of perturbative
 QCD, which are   indicated by red circles  in \fig{amp}-a.  
 In addition, in Ref.\cite{AKLL1} it is shown that  for such $Y$, we can
 safely use the Mueller-Patel-Salam-Iancu (MPSI) approach\cite{MPSI}.
  In this approximation we can use the parton cascade of the
 Balitsky-Kovchegov\cite{BK} equation   to find amplitudes
 $ N\Lb Y- Y', r, \{\vec{r}_i,\vec{b} - \vec{b}_i\}\Rb$ and
   $N\Lb  Y', R, \{\vec{ r}'_i,\vec{b}'_i\}\Rb$, which can be viewed 
as a sum of the BFKL Pomeron 'fan'  diagrams (see \fig{amp}-a for examples of
 such diagrams) . This amplitude can be 
written as  $N\Lb Y- Y', r, \{ r_i,b_i\}\Rb\,\,=\,\,N^{\rm BK}\Lb Y- Y', r,
 \{ r_i,b_i\}\Rb$ (see \fig{amp}-c) with 
    \beq \label{TI3}
N^{\rm BK}\Lb Y- Y', r, \{ r_i,b_i\}\Rb=\sum^{\infty}_{n=1}
 \,\Lb - \,1\Rb^{n+1} \widetilde{C}_n\Lb
  r\Rb \prod^n_{i=1} G_\pom\Lb Y - Y';  r, r_i , b_i\Rb\,\,
=\,\,\sum^{\infty}_{n=1} \,\Lb - \,1\Rb^{n+1} \widetilde{C}_n\Lb
 r\Rb \prod^n_{i=1} G_\pom\Lb z - z_i\Rb
\eeq   
$G_\pom$  denotes the Green function of the BFKL Pomeron. In the last
 equation we used the fact that in the saturation region this Green 
function
 has  geometric scaling behavior, and so it depends on one variable:
 $z_i \,=\,\ln\Lb Q^2_s(Y') r^2_i\Rb$, where $Q_s\Lb Y'\Rb$, is the
 saturation scale,  in the vicinity of the saturation scale\cite{MUTR}
\beq \label{VQS1}
G_\pom\Lb  z_i\Rb\,=\,\phi_0 \Lb r^2_i\,Q^2_s\Lb Y, b_i\Rb\Rb^{1
 - \gamma_{cr}}
~~~~~~\mbox{with} ~~~~~~~~\gamma_{cr}=0.37
\eeq
 We wish to stress that this form for the Green function produces 
screening corrections which increase with increasing energy, and are in 
accord with the behaviour of the high energy LHC results.

  In Ref.\cite{LEPP}, it was shown that, 
 the solution to the non-linear BK equation has   the following general 
form
 \beq \label{TI4}
N\Lb G_\pom\Lb \phi_0,z\Rb\Rb \,\,=\,\,\sum^{\infty}_{n=1} \,\Lb - 
\,1\Rb^{n+1} C_n\Lb \phi_0\Rb G_\pom^n\Lb \phi_0,z\Rb.
\eeq   
Comparing \eq{TI3} with \eq{TI4} we see 
\beq \label{TI5}
\widetilde{C}_n\Lb  r\Rb\,\,\,=\,\,\,C_n\Lb \phi_0\Rb.
\eeq 
Coefficients $C_n$ can be determined from the solution to the 
Balitsky-Kovchegov
 equation \cite{BK}, in the saturation region. The numerical
 solution for the simplified
 BFKL kernel, in which only the leading twist contribution was taken into 
account, is given in Ref. \cite{LEPP}:
 
\beq \label{T16}
N^{\rm BK}\Lb G_\pom\Lb \phi_0,z\Rb\Rb \,\,=\,\,a\,\Lb 1
 - \exp\Lb -  G_\pom\Lb \phi_0,z\Rb\Rb\Rb\,\,+\,\,\Lb 1 - a\Rb
\frac{ G_\pom\Lb \phi_0,z\Rb}{1\,+\, G_\pom\Lb \phi_0,z\Rb},
\eeq
with $a$ = 0.65. \eq{T16} is a convenient parameterization of the
 numerical solution,  having an accuracy of better than 5\%.
Having  determined $C_n$, we can calculate the Green function of the 
dressed BFKL
 Pomeron using \eq{GFMPSI}, and the property of the BFKL Pomeron exchange:
\beq \label{POMTUN}
\frac{\as^2}{4 \pi} \,\,G_\pom\Lb Y - 0, r, R; b  \Rb = \int d^2 r'
  d^2 b' \, d^2 r'' \,d^2 b'' \,G_\pom\Lb Y - Y', r, r', \vec{b}
 - \vec{b}^{\,'} \Rb \, \,
\,G_\pom\Lb Y' r'', R,  \vec{b}^{\,''} \Rb\,\,A^{\rm BA}_{\mbox{\tiny
 dipole-dipole}}\Lb r', r'', \vec{b''} - \vec{b'}\Rb
\eeq 
 
Carrying out the integrations in \eq{GFMPSI}, we obtain the Green
 function of the dressed Pomeron in the following form:
 
 \beq \label{G}
G^{\mbox{\tiny dressed}}\Lb T\Rb = a^2 (1 - \exp\Lb -T\Rb )  +
 2 a (1 - a)\frac{T}{1 + T} + (1 - a)^2 G\Lb T\Rb ~~\mbox{with}~~G\Lb T\Rb = 1 - \frac{1}{T} \exp\Lb \frac{1}{T}\Rb
 \Gamma\Lb 0, \frac{1}{T}\Rb
\eeq
where $\Gamma\Lb s, z\Rb$ is the upper incomplete gamma function
 (see Ref.\cite{RY} formula {\bf 8.35}), and $T$ denotes  the BFKL Pomeron 
in the
 vicinity of the saturation scale ( see \eq{VQS1})
  \beq \label{T}
T\Lb r_\bot, Y\,=\,\ln\Lb s/s_0\Rb, b\Rb\,\,=\,\,\phi_0  \Lb r^2_\bot Q^2_s\Lb Y, b\Rb\Rb^{\bar
 \gamma}  
\eeq
 
The Green function of \eq{G} depends on the size of the dipoles.
 In our analysis of the
 soft interaction we fixed $r = 1/m$, and $m$ was a fitting parameter.

{\it Phenomenology: assumptions and  new small parameters. } \\
  Due to the embryonic stage of theoretical understanding 
of the confinement of quarks and gluons,  it is necessary to use pure
 phenomenological ideas to  ameliorate two major problems in high 
energy 
scattering: the structure of hadrons, and the large impact parameter 
behavior
 of the scattering amplitude\cite{KOWI}.
   To  correct the large impact parameter
 behaviour,  we assume that the saturation momentum  has the 
following dependence  on the 
impact parameter $b$: 
    \beq \label{QS}
Q^2_s\Lb b, Y\Rb\,\,=\,\,Q^2_{0s}\Lb b, Y_0\Rb\,e^{\lambda \,(Y - Y_0)}
~~\mbox{where}~~
Q^2_{0s}\Lb b, Y_0\Rb\,\,=\,\, \Lb m^2\Rb^{1 - 1/\bar \gamma}\,\Lb S\Lb b,
 m\Rb\Rb^{1/\bar{\gamma}} 
~~S\Lb b , m \Rb \,\,=\,\,\frac{m^2}{2 \pi} e^{
 - m b}~~\mbox{and}~~\bar \gamma\,=\,0.63
\eeq 

 We have introduced a new phenomenological parameter $m$ to
 describe the large $b$ behaviour. The $Y$ dependence  as well
 as  $r^2$ dependence, can be found from CGC/saturation approach 
\cite{KOLEB},
 since $\phi_0$ and $\lambda$ can be calculated in the leading order of
 perturbative QCD. However, since the higher order corrections turn out
 to be large \cite{HOCOR}, we treat them as parameters to be fitted. $m$
 is a non-perturbative parameter, which determines the typical sizes of
 dipoles within the hadrons. In Table 1, we show that from the fit,  
 $m$ = 5.25 GeV, supporting our main assumption that we can
 apply the BFKL Pomeron calculus, based on perturbative QCD, to the
 soft interaction since $m \,\gg\,\mu_{soft}$, where $\mu_{soft}$ is
 the scale of soft interaction, which is of the order of the mass of
 pion or $\Lambda_{\rm QCD}$.

 The second unsolved problem for which we need  a phenomenological input,
 is the structure of the scattering hadrons.
 We use a two channel model, which allows us to calculate the
 diffractive production in the region of small masses.
   In this model, we replace the rich structure of the 
 diffractively produced states, by a single  state with the wave 
function 
$\psi_D$, a la Good-Walker \cite{GW}.
  The observed physical 
hadronic and diffractive states are written in the form 
\beq \label{MF1}
\psi_h\,=\,\alpha\,\Psi_1+\beta\,\Psi_2\,;\,\,\,\,\,\,\,\,\,\,
\psi_D\,=\,-\beta\,\Psi_1+\alpha \,\Psi_2;~~~~~~~~~
\mbox{where}~~~~~~~ \alpha^2+\beta^2\,=\,1;
\eeq 

Functions $\psi_1$ and $\psi_2$  form a  
complete set of orthogonal
functions $\{ \psi_i \}$ which diagonalize the
interaction matrix $T$
\beq \label{GT1}
A^{i'k'}_{i,k}=<\psi_i\,\psi_k|\mathbf{T}|\psi_{i'}\,\psi_{k'}>=
A_{i,k}\,\delta_{i,i'}\,\delta_{k,k'}.
\eeq
The unitarity constraints take  the form
\beq \label{UNIT}
2\,\mbox{Im}\,A_{i,k}\left(s,b\right)=|A_{i,k}\left(s,b\right)|^2
+G^{in}_{i,k}(s,b),
\eeq
where $G^{in}_{i,k}$ denotes the contribution of all non 
diffractive inelastic processes,
i.e. it is the summed probability for these final states to be
produced in the scattering of a state $i$ off a state $k$. In \eq{UNIT} 
$\sqrt{s}=W$ denotes the energy of the colliding hadrons, and $b$ 
the impact  parameter.
A simple solution to \eq{UNIT} at high energies, has the eikonal form 
with an arbitrary opacity $\Omega_{ik}$, where the real 
part of the amplitude is much smaller than the imaginary part.
\beq \label{A}
A_{i,k}(s,b)=i \Lb 1 -\exp\Lb - \Omega_{i,k}(s,b)\Rb\Rb,~~~~~~~
G^{in}_{i,k}(s,b)=1-\exp\Lb - 2\,\Omega_{i,k}(s,b)\Rb.
\eeq
\eq{A} implies that $P^S_{i,k}=\exp \Lb - 2\,\Omega_{i,k}(s,b) \Rb$, is 
the probability that the initial projectiles
$(i,k)$  reach the final state interaction unchanged, regardless of 
the initial state re-scatterings.
\par

 The first approach is to use the eikonal approximation for $\Omega$ in which
 \beq \label{EAPR}
 \Omega_{i,k}(r_\bot, Y - Y_0,b)\,\,=\,\int d^2 b'\,d^2 b''\,
 g_i\Lb \vec{b}',m_i\Rb \,G^{\mbox{\tiny dressed}}\Lb T\Lb r_\bot,
 Y - Y_0, \vec{b}''\Rb\Rb\,g_k\Lb \vec{b} - \vec{b}'\ - \vec{b}'',m_k\Rb 
 \eeq 
 where $m_i$ denote the masses, which is introduced phenomenologically to
 determine the $b$ dependence of $g_i$ (see below).
 
 We propose a more general approach, which takes into account the new
 small parameters, that are determined by fitting to the experimental 
data
 (see Table 1 and \fig{amp} for notation):
 \beq \label{NEWSP}
 G_{3\pom}\Big{/} g_i(b = 0 )\,\ll\,\,1;~~~~~~~~ m\,\gg\, m_1 
~\mbox{and}~m_2
 \eeq
 
 The second equation in \eq{NEWSP} leads to the fact that $b''$ in 
\eq{EAPR} is much
 smaller than $b$ and $ b'$,
  therefore, \eq{EAPR} can be re-written in
 a simpler form
 \beq \label{EAPR1}
 \Omega_{i,k}(r_\bot, Y - Y_0, b)\,=\,\underbrace{\Bigg(\int d^2 b''\,
G^{\mbox{\tiny dressed}}\Lb
 T\Lb r_\bot, Y - Y_0, \vec{b}''\Rb\Rb\Bigg)}_{\tilde{G}^{\mbox{\tiny dressed}}\Lb r_\bot, Y - Y_0\Rb}\,\int d^2 b' g_i\Lb
 \vec{b}'\Rb \,g_k\Lb
 \vec{b} - \vec{b}'\Rb \eeq

Using the first small parameter of \eq{NEWSP}, we  see 
 that the main contribution stems from the net diagrams shown in \fig{amp}-b.
 The sum of these diagrams\cite{GLM2CH} leads to the following expression 
for $
 \Omega_{i,k}(s,b)$
 \beq \label{OMEGA}
\Omega\Lb r,  Y-Y_0; b\Rb~=~ \int d^2 b'\,
\,\,\,\frac{ g_i\Lb\vec {b}'\Rb\,g_k\Lb\vec{b} -
 \vec{b}'\Rb\,\tilde{G}^{\mbox{\tiny dressed}}\Lb r, Y - Y_0\Rb
}
{1\,+\,G_{3\pom}\,\tilde{G}^{\mbox{\tiny dressed}}\Lb r, Y - Y_0\Rb\left[
g_i\Lb\vec{b}'\Rb + g_k\Lb\vec{b} - \vec{b}'\Rb\right]} ;~~~
g_i\Lb b \Rb\,=\,g_i \,S_p\Lb b; m_i \Rb ;
\eeq
where
\beq \label{SB}
S_p\Lb b,m_i\Rb\,=\,\frac{1}{4 \pi} m^3_i \,b \,K_1\Lb m_i b
 \Rb~~~\xrightarrow{\mbox{Fourier image} }~~~\frac{1}{\Lb 1
 + Q^2_T/m^2_i\Rb^2};~~
\tilde{G}^{\mbox{\tiny dressed}}\Lb r, Y -Y_0\Rb\,\,=\,\,\int d^2 b
 \,\,G^{\mbox{\tiny dressed}}\Lb T\Lb r, Y - Y_0, b\Rb\Rb
 \eeq
where $ T\Lb r, Y - Y_0, b\Rb$ is given by \eq{T}.

The impact parameter  dependence  of $S_p\Lb b,m_i\Rb$ is purely 
phenomenological,
 however, \eq{SB} which has a form of the electromagnetic proton  form 
factor,
 leads to the correct ($\exp\Lb - \mu b\Rb$) behavior at large 
$b$\cite{FROI},
 and has correct behavior at large $Q_T$, which has been calculate in the
 framework of perturbative QCD \cite{BRLE}. We wish to draw the reader's 
attention to the fact 
 that $m_1$ and $m_2$ are the two dimensional scales in a hadron, which in
 the framework of the constituent quark model,  we assign to the size of 
the
 hadron ($R_h \propto 1/m_1$), and the size of the constituent quark
 ($R_Q 
\propto 1/m_2$).
Note  that  $\tilde{G}^{\mbox{\tiny dressed}}\Lb Y - Y_0\Rb$ does not 
depend
 on $b$.  In all previous formulae, the value of the triple BFKL Pomeron
 vertex
 is known: $G_{3 \pom} = 1.29\,GeV^{-1}$.
  
\begin{table}[h]
\begin{tabular}{|l|l|l|l|l|l|l|l|l|l|}
\hline
model &$\lambda $ & $\phi_0$ ($GeV^{-2}$) &$g_1$ ($GeV^{-1}$)&$g_2$
 ($GeV^{-1}$)& $m(GeV)$ &$m_1(GeV)$& $m_2(GeV)$ & $\beta$ &$\chi^2$/d.o.f.\\
\hline
I(soft int.)& 0.38& 0.0019 & 110.2&  11.2 & 5.25&0.92& 1.9 & 0.58 &1.02 \\
\hline
II:(soft + DIS)& 0.38& 0.0022 & 96.9&  20.96 & 5.25&0.86& 1.76 & 0.66&1.28
  \\
 \hline 
  
  \end{tabular}
\caption{Fitted parameters of the model. Fit I: parameters for the
 soft interaction at high energy are  taken 
from Ref.\cite{GLM2CH}. The additional parameters for DIS were
 found by fitting to the $F_2$ structure function\cite{GLP}.
 Fit II: joint fit to the soft interaction data  
 at high energy and 
the DIS data. The values of $\chi^2/d.o.f.$ were calculated for the
 published data points at energies W = 0.546, 1.8, 7 an 8 TeV
 shown in \fig{fit}.}  
\label{t1}
\end{table}

\begin{figure}[ht]
    \centering
  \leavevmode
  \begin{tabular}{c c c}
      \includegraphics[width=6cm,height=5cm]{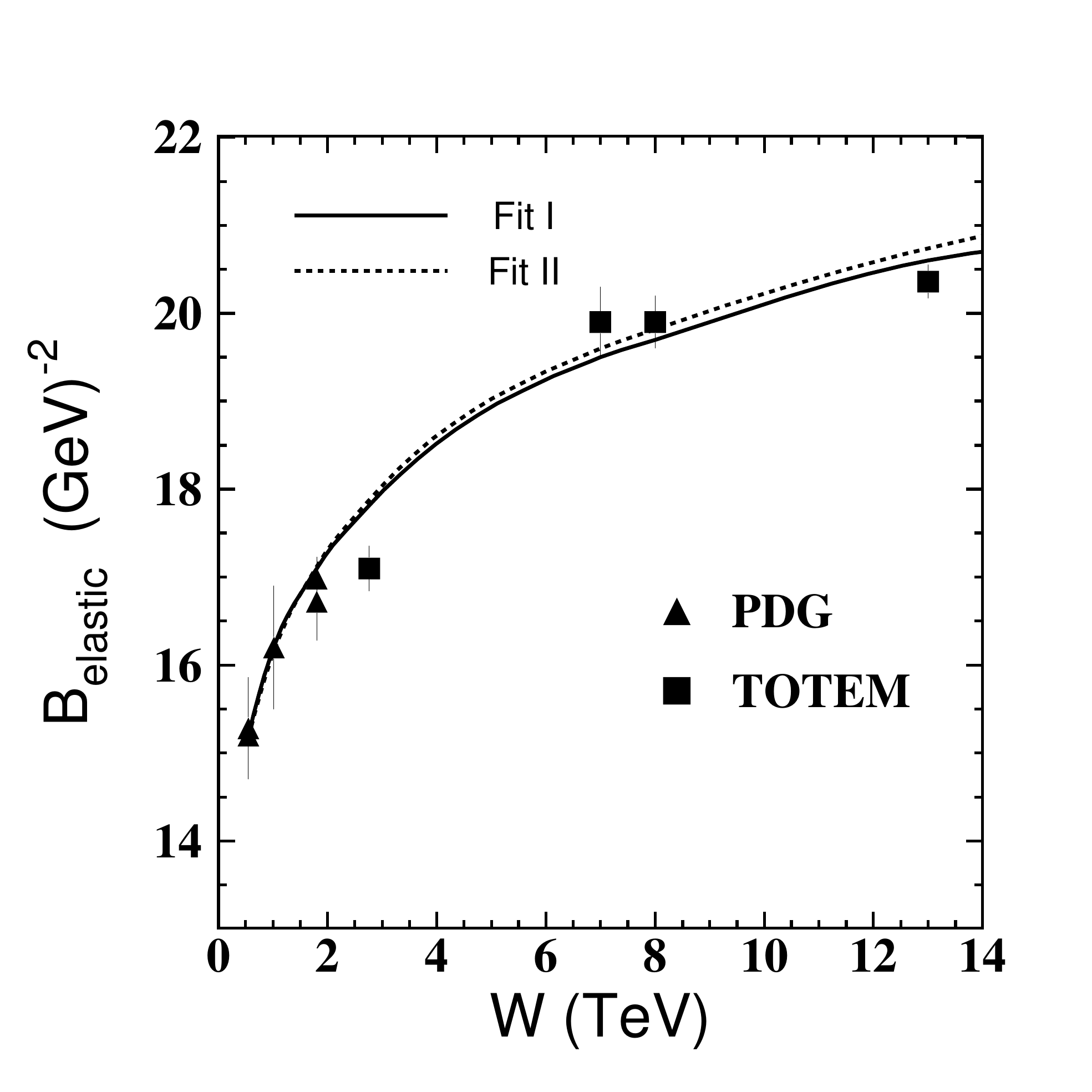}  &\includegraphics[width=6cm,height=5cm]{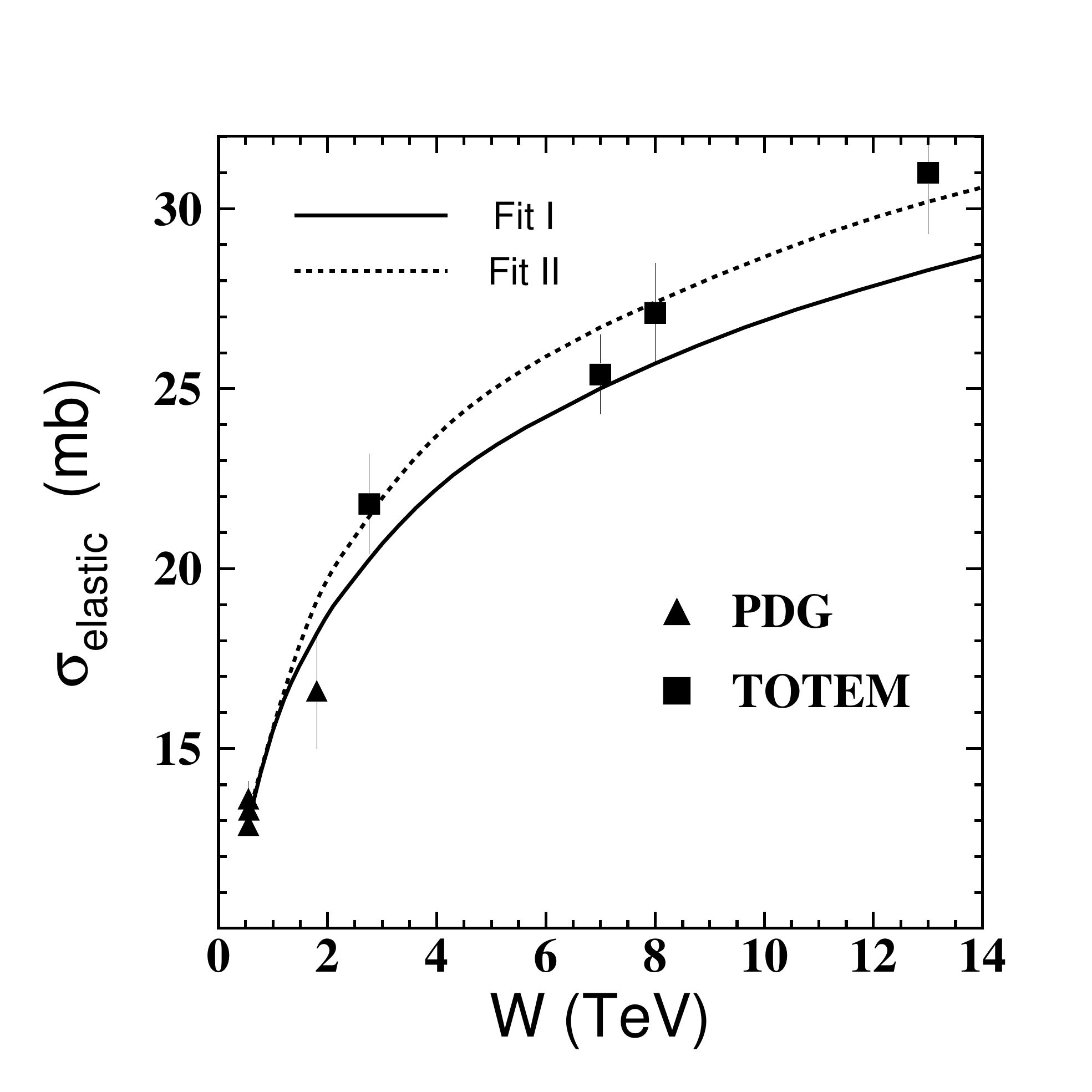}&\includegraphics[width=6cm,height=5cm]{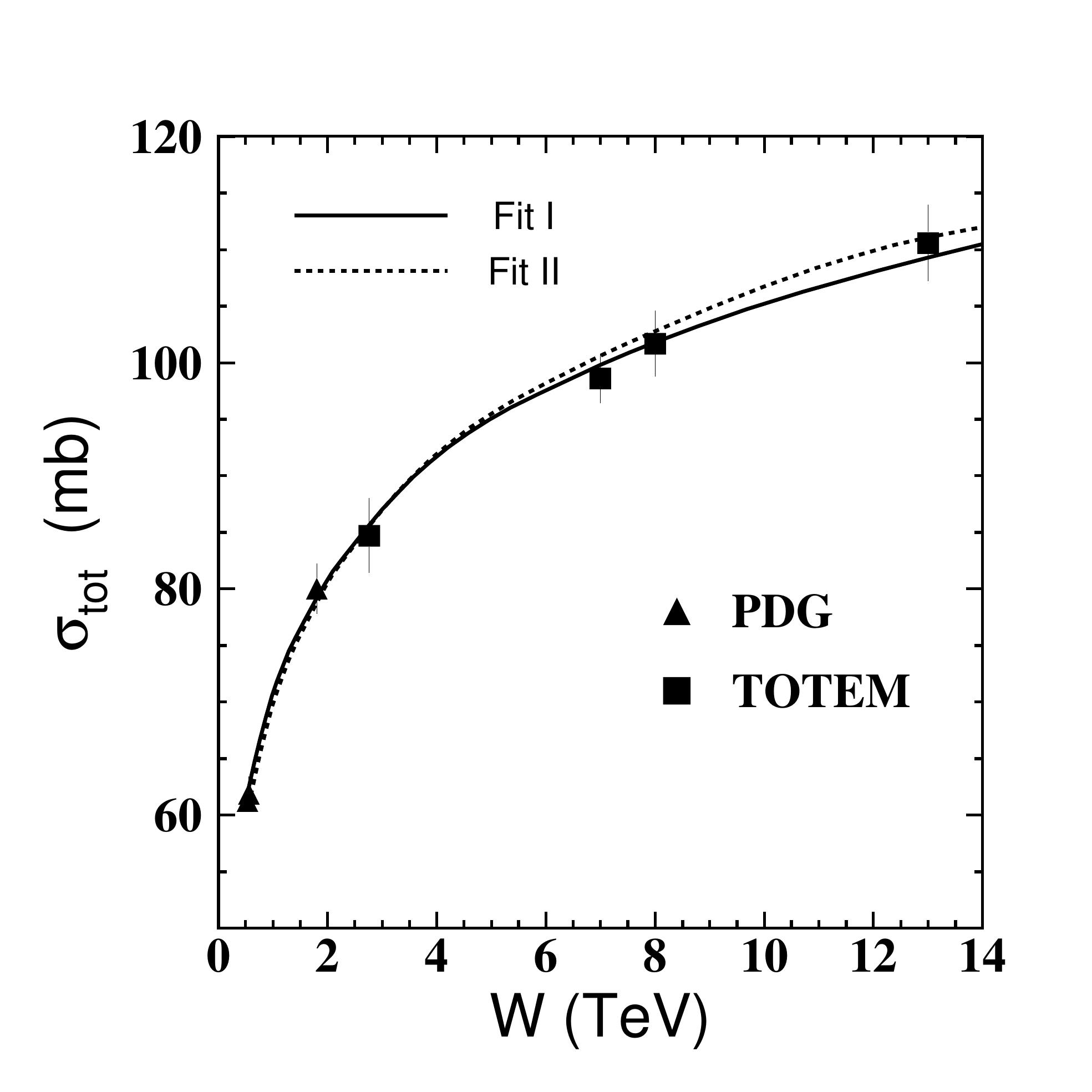} \\
      \fig{fit}-a&\fig{fit}-b& \fig{fit}-c\\
      \end{tabular}
      
 \caption{ The energy behaviour of $\sigma_{tot},\sigma_{el}$ and the
 slope $B_{el}$  in our model. Fit I and fit II are explained in the text.
 Data are taken from Refs.\cite{PDG,TOT,Csorgo}   }
 \label{fit}
   \end{figure}

 { \it Results of the fit.}\\In this paper we make two fits. In the first
 one (fit I in Table 1) we do not change the parameters that govern the
 soft interactions in 
our
 model, and  are shown in Table 1 and \fig{fit}. The additional parameters
 that we 
need
 for  the description of the deep inelastic data \cite{GLP}  were fitted
 using the HERA data on the deep
 inelastic structure function $F_2$. The second fit, is a joint fit to 
the
 soft strong interaction data and the DIS data.
 The model predictions 
are in accord with the data for $0.85 \leq Q^{2} \leq 27 GeV^{2}$, while 
for 
higher values of $Q^{2}$ and of $x$, the model values
 are slightly larger than the data(see Ref.\cite{GLP}).


 In Table 2 we present 
our
 predictions for the soft interaction observables,  
in general the values obtained in the model for the soft interactions
agree with the published  LHC data, as well as the new preliminary TOTEM 
values at W = 2.7 and 13 TeV (see Ref.\cite{TOT,Csorgo}).
We are in very good agreement with the data for 
$\sigma_{tot}$, $\sigma_{el}$ and $B_{el}$.

 Since the TOTEM prelimenary results for energies W = 2.7 and 13 TeV, 
were only released after our paper \cite{GLP} was published, the numbers 
appearing in Table II for these energies can be considered as predictions 
of our model, and are in good agreement with the TOTEM data. The one 
outstanding discrepancy is the value for $B_{elastic}$ at W = 2.76 TeV, 
where our result of 17.8 $GeV^{-2}$ is higher than the TOTEM experimental 
value of 17.1 $\pm$ 0.26 $GeV^{-2}$, however,  it is a smooth function 
of energy, and  does not exhibit  a break in 
the energy  dependence of $B_{elastic}$, as was proposed  by 
Cs$\ddot{o}$rg\H{o}\cite{Csorgo}.

 Regarding $\sigma_{sd}$ and 
$\sigma_{dd}$, a problem exists when attempting to compare with the 
experimental results.
This is due to the difficulties of measuring diffractive events at 
LHC energies, the different experiments have  different cuts on the 
values of the diffractive mass measured, making it problematic when 
attempting to compare the model predictions with the experimental results.

\begin{center}
\begin{table}
\begin{tabular}{|l|l|l|l| l l |l l|}
\hline
W &$\sigma_{tot} $& $\sigma_{el}$(mb) &$B_{el}$&~~~single& diffraction~~ &~~~~double& diffraction~~~ \\
(TeV)   &  (mb)  &        (mb)              &          $(GeV^{-2})$& $\sigma^{\rm smd}_{\rm sd}$ (mb)  &$\sigma^{\rm lmd}_{\rm sd}$ (mb)& $\sigma^{\rm smd}_{\rm dd}$ (mb)&$\sigma^{\rm lmd}_{\rm dd}$ (mb)\\
\hline
0.576 &62.3(60.7)& 12.9(13.1)&15.2(15.17) & 5.64(4.12)& 1.85(1.79)& 0.7(0.39)&0.46 (0.50)\\
\hline
0.9 & 69.2(68.07) &15(15.05)&16(15.95) &6.254.67)& 2.39(2.35)& 0.77(0.46)&0.67(0.745) \\
\hline
1.8&79.2(78.76)&18.2(19.1)&17.1(17.12)&7.1(5.44)&3.35(3.28) & 0.89(0.56)&1.17 (1.30) \\
\hline
2.74 &85.5(85.44)&20.2(21.4)&17.8(17.86)&7.6(5.91)&4.07(4.02) &0.97(0.63)&1.62(1.79)\\
\hline
7 &99.8(100.64)&25(26.7)&19.5(19.6)&8.7(6.96)& 6.2(6.17)&1.15(0.814)&3.27(3.67)\\
\hline
8 & 101.8(102.8)&25.7(27.4)&19.7(19.82)&8.82(7.1)&6.55(6.56) &1.17(0.841)&3.63(4.05)\\
\hline
13 & 109.3(111.07)&28.3(30.2)&20.6(20.74)&9.36(7.64)& 8.08(8.11) & 1.27(0.942)&5.11(5.74)\\
\hline
14 & 110.5(111.97)&28.7(30.6)&20.7(20.88)&9.44(7.71)& 8.34(8.42) & 1.27(0.96) &5.4(6.06)\\
\hline
57 & 131.7(134.0)&36.2(38.5) &23.1(23.0)&10.85(9.15)&15.02(15.01) & 1.56(1.26) &13.7(15.6)\\
\hline
\end{tabular}
\caption{ The values of cross sections and elastic slope versus
 energy. $\sigma^{\rm smd}_{\rm sd}$  and $\sigma^{\rm smd }_{\rm dd}$
 denote the cross sections for  diffraction dissociation
 in the small mass region, for single and double diffraction, which stem
 from the Good-Walker mechanism. While  $\sigma^{\rm lmd}_{\rm sd}$  and 
$\sigma^{\rm lmd}_{\rm dd}$
 denote high mass  diffraction, coming from the dressed 
Pomeron
 contributions.  The predictions of  fit II,  are shown in brackets.}
\label{t2}
\end{table}
\end{center}
In Table 2 we show the results of the two fits, the results  
are close to one another,
  the main difference shows up only at high energies. Indeed, in fit I 
the
 cross section for single diffraction is equal to 14.9 mb, while in fit II
 this value is smaller (13.1 mb).  The smaller value of the
 diffraction cross sections is closer to TOTEM and CMS data.
 
 {\it Conclusions.}\\ 
\fig{fit} shows that we are able to describe the
 experimental data on soft interaction at high energies including the
 LHC data.  \fig{fit} shows that the screening corrections become
 stronger at higher energies.
 Our model also  describes the wide
 range of the experimental observables: the DIS data from HERA, the 
   total,
 inelastic, elastic and diffractive cross sections, the $t$-dependence of
 these cross sections as well as the inclusive production and rapidity
 and angular
 correlations in the wide range of energies including the LHC data.
 We believe that this fact is a strong argument in favour of the 
CGC/saturation  approach.

  {\it Acknowledgements.} \\
   We thank our colleagues at Tel Aviv University and UTFSM for
 encouraging discussions.
 This research was supported by the BSF grant   2012124, by 
   Proyecto Basal FB 0821(Chile) ,  Fondecyt (Chile) grant  
 1140842, and by   CONICYT grant PIA ACT1406.

\end{document}